\documentclass[prl,twocolumn,superscriptaddress,showpacs,floatfix]{revtex4}

\usepackage{graphicx}
\usepackage{bm}

\usepackage{hyperref}
\hypersetup{
breaklinks=true,
hyperfootnotes= true,
pagecolor=white,
colorlinks=true,
linkcolor= black,
citecolor= black,
urlcolor=blue
}
\begin{document}
\title{Quantifying turbulence induced segregation of inertial particles}

\author{Enrico Calzavarini}
\altaffiliation{present address: \'Ecole Normale Sup\'erieure de Lyon, CNRS UMR5672, 46 All\'ee d' Italie, 69007 Lyon, France.}
 \affiliation{
Department of Applied Physics, JMBC Burgers Center for Fluid Dynamics,
and IMPACT Institute, University of Twente, P.O Box 217, 7500 AE Enschede, The Netherlands
}

\author{Massimo Cencini} \affiliation{INFM-CNR, SMC Dept.\ of Physics,
  Universit\`a ``La Sapienza'', P.zzle A.~Moro~2, 00185 Roma,
  \ and\\ CNR-ISC, Via dei Taurini 19, 00185 Roma, Italy}

\author{Detlef Lohse} \affiliation{
Department of Applied Physics, JMBC Burgers Center for Fluid Dynamics,
and IMPACT Institute, University of Twente, P.O Box 217, 7500 AE Enschede, The Netherlands
}

\author{Federico Toschi} \affiliation{CNR-IAC, Viale del Policlinico 137, 00161 Roma,
  \ and\\ INFN, Sezione di Ferrara, via G.\ Saragat 1, 44100 Ferrara, Italy}

\collaboration{International Collaboration for Turbulence Research}
\noaffiliation

\begin{abstract}
Particles with density different
from that of the advecting fluid
cluster due to the different response of light/heavy particles
to turbulent fluctuations. This study focuses on the quantitative
characterization of the {\em segregation} of dilute poly-disperse inertial particles evolving in
turbulent flow, as obtained from Direct Numerical
Simulation of homogeneous isotropic turbulence. We introduce an indicator of segregation
amongst particles of different inertia and/or size, from which a length scale
$r_{seg}$, quantifying the segregation degree between two particle types, is deduced.
\end{abstract}
\pacs{47.27.-i, 47.10.-g}
\date{\today}

\maketitle 

The ability of efficiently mixing transported substances is one of the
most distinctive properties of turbulence, which is ubiquitous in
geophysical and astrophysical fluids. New features appear when
turbulent flows are seeded with finite-size particulate matter having
density $\rho_p$ different from the carrier fluid density
$\rho_f$. Due to inertia, measured by the Stokes time
$\tau\!\!=\!\!a^2/(3\beta \nu)$ ($a$ being the particle radius and
$\nu$ the fluid viscosity; $\beta\!=\!3\rho_f/(2\rho_p\!+\!\rho_f)$),
such particles detach from fluid parcels' paths and distribute
inhomogeneously~\cite{m87,cfptv92,bff01}.  Although this phenomenon of
{\em preferential concentration} \cite{ef94} has been known for a long
time~\cite{m87,cfptv92}, it continues to attract much attention (see
\cite{bff01,bec03x,bec:2005,wmb06,agcbw06,prl,maz03b,cal08} and ref.\
therein). It is important for drag reduction by
microbubbles~\cite{bubbles}, for the effects of microbubbles on the
small scales of turbulence~\cite{bubbles1}, for aerosol physics which
is critical for climatological models~\cite{climate}, or to understand
the patchiness of chemical and biological agents in the
oceans~\cite{planckton}.  The key issue is the tendency of inertial
particles to form clusters with the consequent enhancement of the
particle interaction rate.

\begin{figure}[t]
  \includegraphics[width=\hsize]{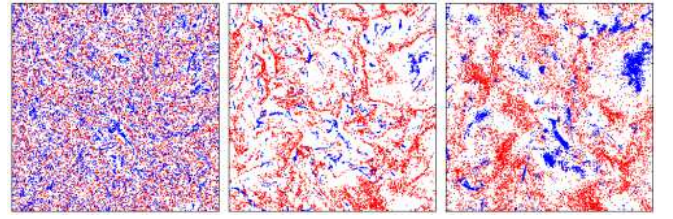}
  \caption{\label{fig:1} Slice 400$\eta\times$400$\eta\times$10$\eta$
of heavy $\beta\!\!=\!\!0$ (red) and light $\beta\!\!=\!\!3$ (blue)
particle positions.  From left to right $St\!=\!\ 0.1, 1, 4.1 $. Data
refer to the simulation at $\mathrm{Re}_\lambda\!=\!180$.}
\vspace{-0.4cm}
\end{figure}

When having {\it different} particle types in the same flow
(polydispersity), the respective particles probe different flow
structures: light particles ($\beta\!>\!1$, e.g., air bubbles in
water) preferentially concentrate in high vorticity regions, while
heavier ones ($\beta\!<\!1$, e.g.\ sand grains in water) are expelled
by rotating regions.  This leads to a {\em segregation} of the
different particle types, which intuitively is characterized by some
{\em segregation length scale}.  An example of particle segregation is
shown in Fig.~\ref{fig:1}, where snapshots of light and heavy
particles' positions are depicted.  The segregation length depends on
both the respective particle densities and Stokes numbers
$St\!=\!\tau/\tau_\eta$, which measure the particle response time
$\tau$ in units of the Kolmogorov time $\tau_\eta$ (characterizing the
smallest active time scale of turbulence).  This Letter aims to
systematically {\it quantify} the segregation length as a function of
both the relative density ($\beta$) and the Stokes number, which
together characterize the particle classes.

To demonstrate our method, we consider a model of
passively advected dilute (to neglect collisions) suspensions of
particles in homogeneous, isotropic turbulence. Particles are
described as material points which are displaced both by inertial
forces (pressure-gradient force, added mass) and viscous forces
(Stokes drag). Additional physical effects - as lift force, history
force, buoyancy or finite-size and finite-Reynolds corrections - which
may becomes important for the cases of light and/or large particles
($\beta St \geq 1$), are here neglected for simplicity.

The particle dynamics
then reads~\cite{mr83,auton} (see also~\cite{bcpp00})
\begin{equation}
{\rm d}_t\bm x=\bm v\,,\qquad {\rm d}_t \bm v=\beta \, {\rm D}_t \bm u +
\tau^{-1}(\bm u -\bm v)\,,
\label{eq:dynamics}
\end{equation}
where $\bm x\,,\bm v$ denote the particle position and velocity,
respectively and $d_t=\partial_t +\bm v\cdot \bm \nabla$ the time
derivative along the particle path.    The incompressible fluid
velocity $\bm u$ evolves according to the Navier-Stokes equations
\begin{equation} {\rm D}_t \bm u= \partial_t \bm u + \bm u \cdot \bm
  \nabla \bm u = -\bm \nabla p/\rho_f +\nu \Delta \bm u + \bm f
  \,,\label{eq:ns}
\end{equation}
where $p$ denotes the pressure and $\bm f$ an external forcing
injecting energy at a rate $\varepsilon=\langle \bm u\cdot \bm
f\rangle$.
Eq.~(\ref{eq:ns}) is evolved by means of a $2/3$-dealiased
pseudospectral code with a second order Adams-Bashforth time
integrator.
The fluid velocity at particle position is
evaluated by means of a three-linear interpolation.
Simulations have been performed in a cubic box of
side $L\!=\!2\pi$ with periodic boundary conditions, and by using
$N^3\!=\!128^3$ and $512^3$ mesh points (reaching Taylor Reynolds
numbers $\mathrm{Re}_\lambda\!=\!75$ and $180$).
The respective Taylor length scales $\lambda \equiv \sqrt{\langle u_x^2\rangle/\langle \partial_x u_x^2\rangle}$ are $\lambda = 13\eta$ and
$21\eta$.
The parameter space $\beta\times St\!\in\![0\!\!:\!\!3]
\times [0\!\!:\!\!4]$ is sampled with $504$ $(\beta,St)$-points with
$N=10^5$ particles per type in the former case and $64$
(optimally chosen by means of a MonteCarlo allocation scheme based on lower
resolution results) with $N=1.6 \cdot 10^6$ in the latter.
Given the small $Re_{\lambda}$ dependence, we will report here mostly results from $\mathrm{Re}_\lambda\!=\!75$
as for that case we have a more complete sampling of the parameter space ($\beta,St$).

A requirement for any segregation indicator is to result in
zero segregation length
for any two statistically independent distributions of
particles coming from the same class of particles,
at least in the limit of infinitely many particles.
If  the observation scale is too small and the number of particle finite, even
independent particle realizations of the same class of particles
 will artificially appear to be segregated.
Therefore, the definition of segregation strictly requires to indicate the observation scale $r$, and it will be sensitive to the particle number.

However, we aim at a robust observable. Classical and natural
observables, as e.g. the minimal distance between different type of
particles strongly depends on particle number and hence are not
robust.  Harmonic averages of particle distances could be sensitive to
small scales and not be spoiled by the large scales, but the choice of
the weight exponent is rather arbitrary.  The use of a density
correlation function $\langle \rho_1(x)\rho_2(x+r) \rangle$
\cite{falco}, though possible, requires to introduce a coarse-graining
scale (to define the densities) which may quantitatively affect the
estimate. The mixed pair correlation function, or mixed radial
distribution function \cite{Zhou01}, is not bounded at small-scales
for clustered distributions of point-particles.

Our approach is inspired by Kolmogorov's distance measure between two
distributions \cite{kolmo63} and is based on particle densities
coarse-grained over a scale $r$, which can be understood as resolution
of a magnifying glass used to look at the segregation. The whole
volume $L^3$ is partitioned into $\mathcal{M}(r)\!=\!  (L/r)^3$
cubes. We then define the following segregation indicator:
\begin{equation}
S_{\alpha_1,\alpha_2}(r)= \frac{1}{N_{\alpha_1}+N_{\alpha_2}}
\sum^{\mathcal{M}(r)}_{i=1} \left| n_i^{\alpha_1} - n_i^{\alpha_2}\right|\,.
\label{eq:obs}
\end{equation}
The subscripts $\alpha_1$ and $\alpha_2$ index the particle
parameters, i.e., $\alpha_1\!=\!(\beta_1,St_1)$ and
$\alpha_2\!=\!(\beta_2,St_2)$, $N_{\alpha}$ is the total number of
particles of $\alpha$-type, while $n_i^{\alpha}$ that of particles contained in
each cube $i$.  The case $\alpha_1\!=\!\alpha_2$ should be considered
as taking independent realizations of the particle distribution (otherwise
$S_{\alpha_1,\alpha_1}\!\equiv\!0$ trivially) so that
$S_{\alpha,\alpha}$
sets the minimum detectable segregation degree.
\begin{figure}[t!]
  \includegraphics[width=0.9\hsize]{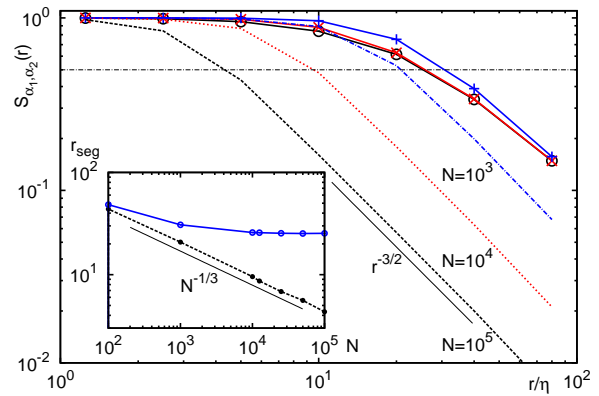}
  \caption{\label{fig:2}$S_{\alpha_1,\alpha_2}(r)$ vs. $r$ for
    $\alpha_1\!=(\beta_1\!=\!0, St_1\!=\!1.1)$ (heavy type) and
    $\alpha_2\!=\!(\beta_2\!=\!3, St_2\!=\!1.1)$ (light type) for
    different particle numbers $N\;
    (=\!N_{\alpha_1}\!=\!N_{\alpha_2})$: $(\circ)$ $N=10^5$,
    $(\times)$ $N=10^4$ and $(+)$ $N=10^3$.  Dashed and dotted lines
    refer to homogeneously distributed particles samples at various
    $N$. The expected Poisson scaling behavior,
    $S_{\alpha_1,\alpha_2}\propto r^{-3/2}$, is also reported.  Inset:
    $r_{seg}$, defined by $S_{\alpha_1,\alpha_2}(r_{seg})=1/2$, as a
    function of $N$ for both heavy vs.\ light particles case and
    Poissonian samples.  While for the former $r_{seg}$ saturates as
    $N$ increases, for the latter $r_{seg}$ goes to zero like
    $r_{seg,h}\propto N^{-1/3}$, as expected for Possonian samples
    (see text for details).}
\vspace{-0.4cm}
\end{figure}

Let us first discuss the limiting cases of
$S_{\alpha_{1},\alpha_{2}}(r)$. First, it can vary in the range
$[0,1]$.  $S_{\alpha_{1},\alpha_{2}}(r)=1$ means that the two
distributions are not overlapping when looked at resolution $r$.  For
small enough scales, i.e., $r \ll 1/\rho^{1/3}$ (which is the mean
distance of two particles with $\rho=N/L^3$ the particle number
density) this holds for any realization, therefore $\lim_{r \to 0}
S_{\alpha_{1},\alpha_{2}} = 1$.  On the contrary $\lim_{r \to L}
S_{\alpha_{1},\alpha_{2}} = 0$ as the total number of particles of the
two species is globally identical (as assumed here). These limiting
cases are observed in Fig.~\ref{fig:2}. Clearly,
$S_{\alpha_{1},\alpha_{2}}$ is a meaningful indicator of segregation
only if it does not depend too severely on the particle number $N$.
Indeed, in Figure~\ref{fig:2}, $S_{\alpha_1,\alpha_2}$ (computed for
the red and blue distribution of the central panel of
Fig.~\ref{fig:1}) shows only a very weak $N$-dependence at
sufficiently large $N$.  This is in contrast with the behavior of
(\ref{eq:obs}) for two independent and homogeneously distributed
particle realizations (also shown in Fig.~\ref{fig:2}).  The latter
case can be easily understood recognizing that in each box of side
$r$, $n_i$ is a Poisson random variable, so that we can estimate $n_i
\approx \rho r^3 \pm \sqrt{\rho r^3}$, where the two terms come from
the average and the fluctuation contributions, respectively. In
eq. (\ref{eq:obs}) the average cancels and, summing the fluctuations
over all the $(L/r)^3$ cells, one has the order of magnitude estimate
$S(r) \sim (L/r)^3 N^{1/2} (r/L)^{3/2}=N^{1/2} (r/L)^{-3/2}$,
explaining both the observed scaling behavior and the strong
dependence on $N$.

The segregation indicator allows us to extract the desired segregation
length scale $r_{seg}$.  This can be done by fixing an arbitrary
threshold value for $S$; we employed
$S_{\alpha_{1},\alpha_{2}}(r_{seg}) =1/2$ (see Fig.~\ref{fig:2}).
With this definition, as shown in the inset of Fig.~\ref{fig:2}, for
truly segregated (heavy vs.\ light) samples $r_{seg}$ saturates with
increasing $N$.  This does not hold for uniformly distributed
(non-segregated) particles.  In the latter case $r_{seg}$ essentially
coincides with the interparticle distance $r_{seg}=r_{seg,h}\approx
1/\rho^{1/3}=L/N^{1/3}$, as also seen from the inset of
Fig~\ref{fig:2}.  The behavior of $r_{seg}$ encompasses the fact that
for a finite number of particles $N$ a natural cut-off distance exists
(the mean inter-particle distance) although we know theoretically that
$\lim_{N \to \infty} r_{seg}(N) = 0$.  Hence $r_{seg,h}(N)$ can be
interpreted as the accuracy in estimating $r_{seg}$ given a finite
particle number $N$.

\begin{figure}[t!]
   \includegraphics[width=0.9\hsize]{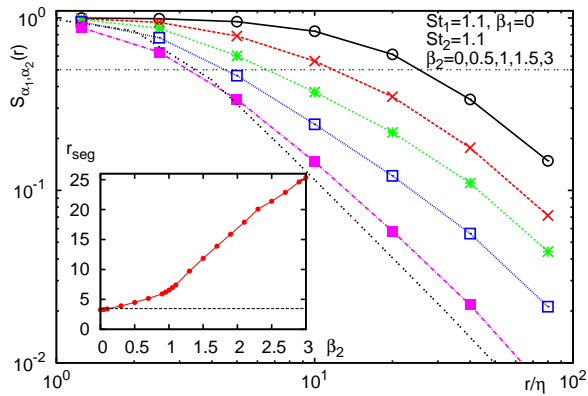}
  \caption{\label{fig:3}
   $S_{\alpha_1,\alpha_2}(r)$ for $\alpha_1\!=\!(\beta_1\!=\!0,
    St_1\!=\!1.1)$ and $\alpha_2\!=\!(\beta_2,St_2\!=\!St_1)$, i.e., a
    heavy particle with $\beta_1=0$ and a given $St$ vs.\ those having the
    same $St$ but different densities $\beta_2$.
    From bottom to top: $\beta_2 = 0, 0.5, 1, 1.5, 3$.
    Inset: $r_{seg}$ vs.\ $\beta_2$, with $r_{seg}$ defined as in Fig~\ref{fig:2}, i.e., $S_{\alpha_1,\alpha_2}(r_{seg})=1/2$. The straight dashed line shows $r_{seg,h} \approx 4\eta \approx 0.3 \lambda$.
}
\vspace{-0.4cm}
\end{figure}

We now calculate $r_{seg}$ for a pair of two different particle classes
to quantify their mutual segregation.
Figure ~\ref{fig:3} displays
$S_{\alpha_1,\alpha_2}(r)$ for distributions composed of heavy
particles with $\beta_1=0$ and $St_1=\mathcal{O}(1)$ and particles
with the same $St(\!=\!St_2\!=\!St_1)$ but different densities. As one
can see, segregation increases with the density difference, but
$r_{seg}\approx r_{seg,h}$ for $\beta_2< 0.5$, meaning that heavy enough
  particles basically all visit the same locations in the flow,
  irrespective of their exact density:
They tend to avoid vortical
regions~\cite{prl}.  A sensitive increase of $r_{seg}$ is observed for
$\beta_2>1$ (Fig.~\ref{fig:3} inset) and as expected the maximal segregation length is obtained
for bubbles, i.e., particles with density ratio $\beta=3$, where
$r_{seg}\approx 25 \eta \approx 1.9 \lambda$.
For the same case, $St=1.1$, $\beta=0$ vs. $\beta=3$, at $Re_{\lambda} = 180$ we find  $r_{seg}\approx 29 \eta \approx 1.4 \lambda$.

\begin{figure}[!t]
\includegraphics[width=1.0\hsize]{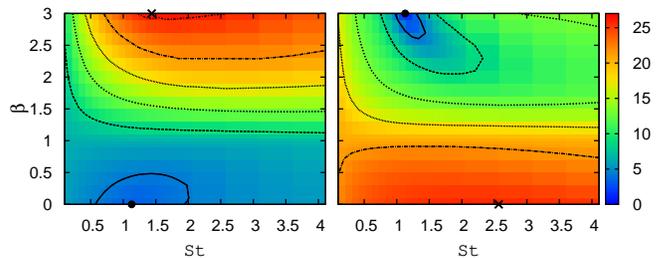}
\caption{\label{fig:4} $r_{seg}$ between particle distributions with
  $\beta\!=\!0$, $St\!=\!1.1$ (left) and $\beta\!=\!3$, $St\!=\!1.1$
  (right) vs. distributions with generic $\beta,St$.  ($\bullet$)
  indicates the reference particle type, and ($\times$) the location
  of the maximal segregation length $r_{seg}^{(max)}$.  The solid
  contour line, traced at $r_{seg}\!=\!r_{seg,h}\!\equiv\!L/N^{1/3}$,
  sets the sensitivity level to distinguish between segregated and
  unsegregated particle distributions.  Dashed and dotted lines are
  drawn at $r_{seg}\!=\!n\cdot r_{seg,h}$, with $n = 2,\ldots,6$.  The
  color scale codes the value of $r_{seg}$ in units of the Kolmogorov
  length, $\eta$.}
  \vspace{-0.4cm}
\end{figure}

Thanks to the large number of particle types in our database we can
extend the study of the segregation length to a wide range of physical
parameters.  In Fig.~\ref{fig:4} we show the value of the segregation
length by fixing $\alpha_1\!=\!(\beta_1,St_1)\!=(0,1.1)$ (left), the
red particles of central panel of Fig.~\ref{fig:1} or the blue ones by
fixing $\alpha_1\!=(3,1.1)$ (right) and varying
$\alpha_2\!=\!(\beta_2,St_2)$ for the second kind of particles. The
emerging picture is as follows.  Particle class pairs with
$St_1\!\approx\! St_2$ and $\beta_1\!\approx \!\beta_2$ have a
segregation length close to the interparticle distance $r_{seg,h}$ and
are unsegregated, while as soon as the Stokes number or the density
difference become larger, $r_{seg}\!>\!r_{seg,h}$. The maximal
segregation length ($r_{seg}^{(max)}\!\approx \!27\eta \!\approx \!
2.1\lambda $) is roughly twice the Taylor microscale and is realized
for particles with large density difference $\beta_1=0$ and
$\beta_2\!>\!1$ (or $\beta_1\!=\!3$ and $\beta_2\!<\!1$). These
results thus confirm those of Fig.~\ref{fig:3}.  It is interesting to
note that heavy couples, $\beta_1,\beta_2\!\leq\! 1$, segregate less
than light ones, $\beta_1,\beta_2\!\geq\! 1$, which are thus much more
sensitive to small variations of density and/or response times. The
correlation between position and flow structure is thus much stronger
for light particles.

\begin{figure*}[!]
 \vspace{-.2 cm}
\includegraphics[width=1\hsize]{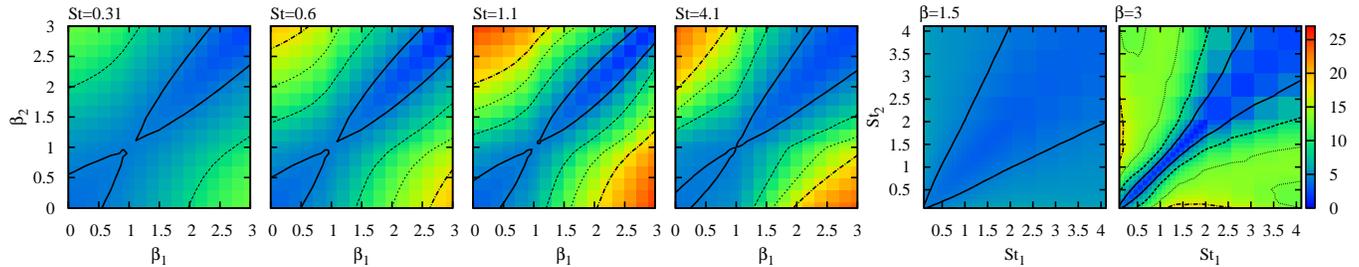}
 \vspace{-.6 cm}
\caption{\label{fig:5} From left to right, segregation length
$r_{seg}$ between
  distribution of particles with $St=(0.31, 0.6, 1.1, 4.1)$ vs. the
  densities $\beta_1,\beta_2$ and (last two panels) for particle
  class pairs having the same densities $\beta_1=\beta_2=1.5$ (resp.\ $3$) and
  different $St$. The color scale and the contour lines are, as in
  Fig.~\ref{fig:4}, at $r_{seg}\!=\!n\cdot r_{seg,h}$, with $n = 2,\ldots,4$. }
  \vspace{-0.4cm}
\end{figure*}

We next systematically study what happens when fixing $St$ or $\beta$.
In the first four panels (from left) of Fig.~\ref{fig:5}, $r_{seg}$ is
shown as a function of $\beta_1$ and $\beta_2$, where we fixed the
Stokes numbers, $St_1\!=\!St_2\!=\!St\!=\!(0.31, 0.6, 1.1, 4.1)$.
Close to the diagonal $\beta_1\approx \beta_2$ it is $r_{seg}\leq
r_{seg,h}$, i.e., similar particles are poorly segregated. Outside
these regions $r_{seg}>r_{seg,h}$ and segregation is above the
accuracy threshold $r_{seg,h}$. Two observations are in order. First,
the strongest segregation is present for the case of $St\!\sim\!1$
meaning that response times of the order of the Kolmogorov time are
best suited to generate strong correlations between flow structures
and particle positions and consequently segregation.  Further,
segregation is stronger for couples composed by very heavy ( $\beta
\sim 0$) and very light ($\beta\sim 3$) particles.  Therefore, light
and heavy particles with $St\sim 1$ display the strongest
clusterization. Second, the numerical value of the segregation length
saturates to a constant value $\approx 2 \lambda$, strongly indicating
that we are measuring an intrinsic property of the underlying
turbulent flow emerging when particles are strongly clusterized.

The two rightmost panels of Fig.~\ref{fig:5} display cuts done by
fixing the density, namely $\beta_1\!=\!\beta_2\!=\!\beta\!=\!1.5$
(resp.\ =3). For $\beta\!<\!1.5$, i.e. relatively heavy or weakly light
particles, there is only a slight tendency toward segregation at
varying the Stokes number.  This means that even if the particles 
form clusters, such clusters are not too sensitive against variation
of $St$.  For very light particles, $\beta\!=\!3$, the situation is
different when comparing the case of $St\!\sim\! 1$ with a $St \sim 0$
case. As expected for $St\! \sim\! 0$, though particles are light, they
distribute almost uniformly (they are not too far from the tracer
limit) while for $St\!\geq\! 1$ they are strongly correlated with the
vortex filaments which are unevenly distributed in the flow~\cite{jim98}.

In conclusion, we introduced an indicator able to
quantify the segregation degree and allowing to define a segregation
length scale $r_{seg}$ between different classes of particles, which
follow simplified dynamical equations in isotropic turbulence.  The
extracted information is in line with the intuitive idea of
expulsion/entrapment of particles due to vortical structures which is
now on a more quantitative ground.  The maximal segregation length,
for instance for heavy particles ($\beta\!=\!0$, $St\!=\!1.1$), is
obtained with bubbles with slightly larger Stokes ($St\! \sim\! 1.4$)
(Fig.~\ref{fig:4} left); it measure $r_{seg}^{(max)}\simeq \!27 \eta
\simeq\! 2.1 \lambda$ at $Re_\lambda\!=\!75$.  At $Re_\lambda=180$ we
get similarly: $r_{seg}^{(max)}\simeq\! 48 \eta \!\simeq \!2.3
\lambda$.  Therefore, $r_{seg}^{(max)}$ is about twice the Taylor
length.

Important areas of application of the developed methods go far beyond
homogeneous isotropic turbulence including, e.g., heterogeneous
catalysis or flotation, where one is interested in the collision
probability of argon bubbles and solid contaminations in turbulent
liquid steel \cite{zha00}.  In this context our finding that
$r_{seg}^{(max)} \simeq 2 \lambda$ suggests that the cleaning
could become less efficient at high $Re_\lambda$, as bubbles and
particles then become more segregated. Hoever, quantitative statement
requires better models for both the flow geometry and the effective
particle force.  Another example for the application of the suggested
method is the formation of rain drops at solid nuclei in
clouds~\cite{falco}, a mechanism which is crucial to develop models
for rain initiation~\cite{wmb06}.  Disregarding particle segregation
in all these examples would lead to estimates of the collision rate
which could be orders of magnitude off.  
 We have already observed that the point-particle model adopted in our study
neglects some hydro-dynamical effects which corresponds to additional
forces in the particle equation. Such forces, when included with
proper modeling, might smooth the intensity of segregation which, as
we have shown, is mainly due to the relative strength of inertial
forces.

Finally, we stress that the introduced segregation
indicator can be employed in all phenomena involving different classes
of segregating objects.  Provided one knows the position of all the
objects, no prior knowledge on the physical mechanism which leads to
segregation is needed.

We thank J. Bec and L. Biferale for fruitful discussions and
R. Pasmanter for prompting Kolmogorov's distance measure to our
attention.  Simulations were performed at CASPUR (Roma, IT) under the
Supercomputing grant (2006) and at SARA (Amsterdam, NL).  Unprocessed
data of this study are publicly available at the
iCFDdatabase~\cite{database}.


\end{document}